\documentclass[prd,
preprint,
superscriptaddress,preprintnumbers,nofootinbib]{revtex4}
\usepackage{graphicx}
\usepackage{epsfig}
\usepackage{bm}
\usepackage{latexsym,amssymb,amsmath,amssymb,wasysym,float}

\usepackage{color}

\usepackage{scrextend}
\usepackage{mathrsfs}
\usepackage{enumitem}

\usepackage[usenames,dvipsnames]{xcolor}
\definecolor{orange}{cmyk}{0,0.5,1,0}
\definecolor{rossoCP3}{cmyk}{0,.88,.77,.40}
\definecolor{graa}{rgb}{0.8,0.8,0.8}
\definecolor{blaa}{rgb}{0.2,0.2,0.6}

\definecolor{darkgreen}{rgb}{0.,0.6,0.}

\begin{document}

\preprint{MPP-2026-27}

\vskip1cm

\title{
{~}\\{~}\\
Moduli Space Quantum Mechanics 
\\{~}\\

}

\author{\bf Luis A. Anchordoqui}

\affiliation{Department of Physics and Astronomy,  Lehman College, City University of
  New York, NY 10468, USA
}

\affiliation{Department of Physics,
 Graduate Center,  City University of
  New York,  NY 10016, USA
}

\affiliation{Department of Astrophysics,
 American Museum of Natural History, NY
 10024, USA
}

\author{\bf Muldrow\nolinebreak~Etheredge}

\affiliation{Max--Planck--Institut f\"ur Physik,  
 Werner--Heisenberg--Institut, 85748 Garching, Germany}

\author{\bf Dieter\nolinebreak~L\"ust}

\affiliation{Max--Planck--Institut f\"ur Physik,  
 Werner--Heisenberg--Institut, 85748 Garching, Germany}

\affiliation{Arnold Sommerfeld Center for Theoretical Physics, 
Ludwig-Maximilians-Universit\"at M\"unchen,
80333 M\"unchen, Germany
}


\begin{abstract}
\noindent {~}\\
In this paper, continuing the discussion about Species Quantum Mechanics, we investigate quantum mechanics in moduli spaces using a mini-superspace approach. From this perspective, moduli-dependent functions can be viewed as operators, and we explore how the taxonomic relations from the Emergent String Conjecture can constrain the non-commutativity between these operators. Next, we study wave functions on moduli spaces, and we find that the geometry of moduli space plays an important role and leads to excited wave functions localised in the bulks of moduli spaces,
 and with positive energy eigenvalues. 
For cases when potentials are present, these effects result in moduli localised away from classical minima, and often result in excited, positive energy  states.
\noindent {~}\\
\end{abstract}

\date{March 2026}
\maketitle


\tableofcontents

\newpage

\section{Introduction
}

Quantisation of gravity is still a challenging enterprise.  String theory is considered the most successful and possibly
unique framework for quantum gravity. During the past years, the 
swampland program~\cite{Vafa:2005ui}
 has provided a new and general perspective on quantum gravity, namely
 addressing the question that explores which effective field theories in the infrared can be consistently embedded in quantum gravity or string theory
in the UV. Moreover, the swampland approach also provided some new insights about certain stability issues, often also denoted as hierarchy problems; e.g. the dark dimension scenario \cite{Montero:2022prj} proposes an
intriguing relation between the smallness of the cosmological constant and the possible existence of large extra dimensions~\cite{Arkani-Hamed:1998jmv,Antoniadis:1998ig}.

Quantum Mechanics is the fundamental physical theory that describes
the behaviour of matter in terms of wave
functions $\psi$, states $|\psi\rangle$  of a Hilbert space $\mathbb
H$, and operators $\hat A$. One basic and well-known principle of quantum mechanics is that, for operators $\hat A$ and $\hat B$, a non-zero value of
the commutator $[\hat A,\hat B]$ sets a fundamental lower limit on the product
of the uncertainties of the two observables~\cite{Robertson:1929zz}. Noncommutative operators do not share a complete set of common eigenfunctions, meaning a system cannot be in a state that has a definite value for both observables at once. For position and momentum operators $\hat x$ and $\hat p$, this is expressed as $[\hat x,\hat p] = i \hbar$,
yielding Heisenberg's uncertainty principle $\Delta x \ \Delta p \geq \hbar/2$~\cite{Heisenberg:1927zz}.

Another important feature of quantum mechanics related to the
uncertainty principle is that some systems, which are classically
unstable, can become stable and well-defined due to quantum effects.
The most well-known example is the 
hydrogen atom, which is described by stable orbits of the electron wave functions, despite having a potential that is classically unstable and unbounded from below.

In string theory, continuous parameters are given by scalar field vevs, and these parametrize a manifold called the moduli space of string vacua. In this work, we investigate quantum mechanics within these moduli spaces. In low-energy effective field theories that come from string theory, there are large numbers of degrees of freedom. This is because at each point in spacetime, the fields fluctuate independently (though they do couple to each other). However, one can often truncate this description by ignoring the spatial variation of the fluctuations, and effectively compactify to a quantum mechanical description in $(0+1)$ spacetime dimensions. 
This truncation complements the ``mini-superspace" formalism of \cite{Misner:1967uu,Wheeler:1968iap,DeWitt:1967yk},
and the reduced description is that of a quantum mechanical particle (potentially with spin) propagating through the space of fields.

In this paper, we extend and generalize the species quantum mechanics work of \cite{Anchordoqui:2025izb} to
explore quantum mechanics in moduli spaces. In particular, we find that the geometry of moduli spaces frequently affects where the wavefunctions are localized. For example, classical potentials often tend to zero in infinite-distance limits (or require a detailed stabilization mechanism to be minimized in the bulk), but we find that the geometry of moduli spaces frequently results in excited moduli wavefunctions that are localized in the bulk, and away from classical minima of the potentials. Furthermore, from the perspective of moduli-space quantum mechanics, many moduli-dependent functions can be viewed as operators, and these operators are generically non-commutative. Under some assumptions, in asymptotic limits of moduli spaces, the commutation relations between these operators are related to the \emph{\textbf{taxonomy rules}} \cite{Etheredge:2024tok} imposed by the Emergent String Conjecture (ESC) \cite{Lee:2019wij}.

One important observation is that quantum gravity effects are expected to become important at a cut-off species scale 
$\Lambda_s$~\cite{Dvali:2007hz,Dvali:2007wp,Dvali:2009ks,Dvali:2012uq}
(for some more recent work 
see~\cite{Castellano:2022bvr,vandeHeisteeg:2022btw,Cribiori:2022nke,Cribiori:2023ffn,Cribiori:2023sch,Calderon-Infante:2023ler,Calderon-Infante:2023uhz,vandeHeisteeg:2023dlw,Castellano:2023aum,Calderon-Infante:2025pls}),
 which can be much lower than the Planck scale $M_p$ whenever one has a large
 number of species becoming light. Concretely the species scale and the number of light species $N_s$ are related as follows~\cite{Dvali:2007hz}
 \begin{equation}
 \Lambda_s =
 N_s^{1/(2-d)}\, ,
 \end{equation}
where $d$ is the number of space-dimensions of the low-energy
effective field theory. The limit of a large number of species  is predicted to
 happen at any perturbative limit of an effective field theory coupled
 to gravity, or equivalently in any infinite distance limit within the
 field space of the quantum gravity completion~\cite{Ooguri:2006in}.
 In fact, the limit of a large number of species coincides with the asymptotic limits in moduli space, where the mentioned taxonomy rules hold.

\newpage
 
Our starting point is the low-energy effective field theory of some
scalar fields (moduli) that are coupled to gravity,\footnote{In this situation, we are studying the bosonic sector only. More supersymmetric theories with fermions will be studied in future work.}
\begin{eqnarray}
S_{\rm EFT}&\sim&\int d^dx \sqrt{-h}\biggl(M_p^{d-2}\bigl( \frac 12R + {1\over 2}g_{ij}(t){\partial_\mu t^i\partial_\mu t^j}+V(t)
+\sum_{n>2}\Lambda_s^{2-n}(t){\cal O}_n(R)\bigr)
\nonumber\\
&+
&
\sum_{n>2}m_t^{d-n}(t)\tilde{\cal O}_n(R)+\dots\biggr)
 \,,\label{eft}
\end{eqnarray}
where $h$ refers to the space-time metric, the scalar fields $t^i$ correspond to the moduli of the string compactification, $g_{ij}(t)$ is the moduli space metric, and $V(t)$ a possible moduli dependent scalar potential. In addition,
this action 
is a double effective field theory (EFT) expansion~\cite{Calderon-Infante:2025ldq} with respect to the two scales that are relevant for our discussion: there is a first set of higher derivate operators ${\cal O}_n(R)$ of dimension $n$ that are determined by the 
moduli dependent species scale $\Lambda_s(t)$~\cite{vandeHeisteeg:2022btw}. 
On the other hand, the second set of higher derivative operators $\tilde {\cal O}_n(R)$
is suppressed by the tower mass scale $m_t(t)$. 

When one considers a dimensional reduction to one dimension ignoring metric fluctuations (i.e. assuming that the metric of the non-compact 4D space is Minkowski), one then obtains a $(0+1)$-dimensional theory
and one finds a sector of the action that is 
\begin{align}
	S\supset \int dx^0 \ \left({1\over 2}g_{ij} \, \dot t^i\dot t^j+ V(t^i) \right) \,,\label{e.moduliworldline}
\end{align}
where the fields $t^i$ are the moduli of the higher-dimensional theory, and $x^0$ is the canonical time of the theory. This action describes  particles with field-space positions and proper time $x^0$.
With this action, one can canonically quantize the position variables $t^i$ and their associated momenta $\pi_i$. Concretely, quantizing the scalars in one dimension, i.e., applying
the rules of quantum mechanics, we obtain
\begin{equation}
\pi_{t,i}={\delta {\cal L}_{\rm EFT}\over \delta(\partial_{x^0} t_i)}
=g_{ij} \dot t^j, \quad \partial_{x^0} t^j=g_{ij}  \dot t^j\, ,\label{momentum}
\end{equation}
where dots refer to $x^0$ differentiation.
Then we impose
the following canonical equal-time commutation relation 
on the pair $(t^i,\pi_{t,j})$ of conjugate variables:
\begin{equation}
\lbrack t^i,\pi_{t,j}\rbrack =\lbrack t^i,g_{jk}\dot t^k\rbrack =i  \delta^i_j\, .
\end{equation}

As observed in \cite{Anchordoqui:2025izb}, dot product relations between moduli-gradients can be re-expressed in terms of commutation relations. In particular, it was observed in \cite{Anchordoqui:2025izb} that the Castellano-Ruiz-Valenzuela (CRV) pattern 
\cite{Castellano:2023stg,Castellano:2023jjt}
between the species scale and the leading tower can be expressed as a commutation relation. This relation led to the concept of {\sl species quantum mechanics}. In this paper, we generalise species quantum mechanics to moduli-space quantum mechanics, which holds everywhere in moduli space (not just asymptotically). Furthermore, we argue that there are also other commutation relations that hold asymptotically. For instance, all of the taxonomy rules of \cite{Etheredge:2024tok} can be expressed as canonical commutation relations.

The organisation of this paper is as follows. In Sec.~\ref{sec:2}, we discuss how dot-product relations between gradients translate into commutation relations. In particular, we argue that the taxonomy rules can asymptotically constrain operator commutation relations. After that
we investigate patterns involving potentials and explore possible cosmological implications.   
In Sec.~\ref{sec:3} we  study wave-functions in moduli-space. We show in examples, both with and without potentials, that the square-integrable wavefunctions have moduli localized in the bulk of moduli-space.\footnote{In this  paper, we will consider only the moduli inherited directly from the $d$-dimensional theory. However, more generally, fermions, other moduli from the compactification, and gravity play a role. Despite this, our statements about wavefunctions being localized in the bulk of moduli due to moduli effects, should be qualitatively unaffected by the presence of additional fermions (though, zero point-energies may shift).}
 This even can happen when the classical potentials do not have minima in the bulk of moduli spaces, due to effects from the geometry of moduli space inducing an effective potential. We place special attention on the two-dimensional hyperbolic plane  with $SL(2,{\mathbb Z})$ modular duality group. In this case modular invariance imposes strong constraints on the operators, like the species scale, as well as on the moduli space wave function.\footnote{For some earlier work on modular invariance in moduli space and in supergravity
see     \cite{Ferrara:1989bc,Ferrara:1989qb,Font:1990gx,Font:1990nt,Cvetic:1991qm}; a recent review is given in \cite{Cribiori:2024qsv}.} As we will show, 
the associated moduli space wave functions are given by the modular invariant
 Eisenstein series: 
 \begin{equation}
 \psi(\tau,\bar \tau)=E_{1/2+i\beta}(\tau,\bar\tau)\, .
 \end{equation}
 We demonstrate that the spectrum of the Laplacian contains, besides plane wave solutions with continuous momenta proportional to $\beta$, an infinite number excited states with a discrete positive energy spectrum and particular
 eigenvalues $\beta_i$,
 where the complex modulus $\tau$ is stabilised in the interior of the moduli space.
 In addition, we compare the moduli space wave functions with the asymptotic form of the species scale, making in this way contact to the taxonomy rules discussed in Sec.~\ref{sec:2}. We also extend the discussion to cases with a non-vanishing potential. Particularly, we show that for potentials, which classically exhibit an unstable run-away behaviour, 
 quantum effects provide an excited, positive energy spectrum, where the
 moduli are  stabilised at some minima of an effective
 potential. Such a changing classical and quantum dynamics bears a
 resemblance to the stable spectrum of the hydrogen atom. Conclusions are drawn in Sec.~\ref{sec:4}.

\section{Commutation relations, patterns,  and taxonomy relations}
\label{sec:2}

Consider a quantum mechanical particle propagating through a moduli space, with an action given by \eqref{e.moduliworldline}. 
The position of the particle in moduli space is denoted by the variable $t$, which corresponds to a scalar field in the effective action.\footnote{Hence $t$ should not be confused with the time variable, which we will denote by
$x^0$.} As discussed in the Introduction, this particle describes part of the mini-superspace description from compactifing a higher-dimensional stringy EFT to (0+1) dimensions.\footnote{Here, for simplicity, we are assuming there are no spinors, additional moduli from the compactification, or additional bosons from the compactification.} This particle, being quantum mechanical, is described by a state in a Hilbert space. On this Hilbert space, various moduli-dependent functions, including the species scale, masses of various other particles, and tensions of branes can be viewed as operators.

What are the commutation relations of such operators? For generic operators at a generic points in moduli space, the commutation relations might not be simple. But, asymptotically in moduli spaces, operators often simplify. This is because commutation relations of operators can be can be recast as dot products between gradients of operators, and, as we will show, dot products of gradients often simplify asymptotically, due to the ESC. As explained in \cite{Anchordoqui:2025izb}, in quantum mechanics the
 commutator between the two operators $F(t)$ and $\dot H(t)$
 can be expressed as
\begin{eqnarray}
\left[ F(t),{d\over dx^0} H(t)\right]=
\partial_i F(t)g^{ij}(t) \partial_j H(t)\, .\label{fhcom5}
\end{eqnarray}
That is,
\begin{eqnarray}
\left[ F(t),\dot H(t)\right]=i~
 \nabla F(t)\cdot \nabla H(t)\, .\label{fhcom}
\end{eqnarray}
Quantum mechanically, this results in a Robertson uncertainty between the operators
\begin{eqnarray}
\Delta F(t)~\Delta\dot H(t)\geq
\frac{1}{2} |\langle\nabla F(t)\cdot \nabla H(t)\rangle|\, .\label{fhuncer}
\end{eqnarray}
Thus, larger dot products result in larger uncertainties.

\subsection{Asymptotic constraints}
We now examine how these dot products for various operators are frequently constrained asymptotically by the ESC. The ESC often\footnote{There is the possibility of ``sliding" \cite{Etheredge:2023odp}, where the dot products are not constant, even asymptotically. In these regimes, the dot products are modified.}
 constrains multiple physically significant moduli-dependent pairs of functions $F(t)$ and $H(t)$ (e.g. masses of towers, tensions of branes, coupling constants, species scale) to asymptotically satisfy relations of the form
\begin{align}
	(\nabla \log F)\cdot (\nabla \log H)\rightarrow \text{constant}\,.\label{e.asymptconst}
\end{align}
These gradients can be defined with respect to the geodesic distance in field space and are often denoted as $\alpha$ vectors, which are computed as
\begin{equation}
\alpha_F=-\nabla \log F\, .
\end{equation}
Asymptotically, at the boundary of moduli space, the $\alpha_F$ are often a constant. In the simple case of one scalar field $t$, 
the distance function is identical  to a  canonically normalised scalar field $\phi$. This field $\phi$ is given in terms of
 a field redefinition of the field $t$ as follows:
\begin{equation}
\phi={\sqrt k}\log t\, .\label{Fieldredef}
\end{equation}
The constant $k$ depends on the specific asymptotic limit under consideration.   For KK modes
corresponding to $p$ compact dimensions, $k$ is given as $k={p+2\over 2p}$, e.g. decompactification from four to six dimensions ($p=2$) corresponds to $k=1$, decompactification from four to five dimensions
and the M-theory limit ($p=1$) correspond to $k=3/2$ and the perturbative string limit corresponds to $k=1/2$ ($p=\infty$).
Then the asymptotic form of $F$ as function of $\phi$  is given  as
\begin{equation}
F(\phi(t))\sim\exp(-\alpha_F\phi)=t^{-\sqrt k\alpha_F}\, .
\end{equation}

The ESC requires infinite-distance limits to be either decompactification limits or perturbative string limits. It follows that 
the ESC constrains the $\alpha_F$, and thus commutation relations, of the following objects in many asymptotic limits of moduli spaces:
\vskip0.3cm
\begin{itemize}[noitemsep,topsep=0pt]
	\item The species scale $\Lambda_s(t)$, or equivalently the species number $N_s(t)$.	The species scale  asymptotically scales as
	\begin{eqnarray}
\Lambda_s(\phi)\sim \exp({-\alpha_s\phi})=t^{-{\sqrt k\alpha_s}}\, .
\end{eqnarray}
In the perturbative string limit $\alpha_s$ is given as
\begin{equation}\alpha_s={1\over \sqrt {d-2}}\, .
\end{equation}
On the other hand in the  decompactification limit from $d$ to $D=d+p$ dimensions the corresponding $\alpha_s$ has the form
\begin{eqnarray}
\alpha_s=\sqrt{p\over  (D-2)(d-2)}\, .\label{decomp}
	\end{eqnarray}

	\vskip0.3cm
	
	\item Tensions of branes $T(t)$ (including masses $m(t)$ of particle towers and instanton actions $S(t)$). For example the lightest mass has the following asymptotic form
	\begin{eqnarray}
m_\text{lightest}(\phi)\sim \exp({-\alpha_m\phi})=t^{-{\alpha_m\over\sqrt2}}\, ,
\end{eqnarray}
	with
	\begin{equation}
	\alpha_m=\sqrt{\frac{d+p-2}{p(d-2)}}\, ,
	\end{equation}
valid both in the decompactification limit and in the perturbative string limit ($p\rightarrow\infty$).
	
	\vskip0.3cm
	\item The determinant of the metric on moduli space $g(t)=\det{g_{ij}(t)}$.
	\vskip0.3cm
	\item Gauge couplings $e(t)$ (including axion decay constants $f(t)$).
\end{itemize}

\vskip0.3cm
The product rules governing the species scale, particle masses, and brane tensions have been derived in \cite{Castellano:2023jjt, Etheredge:2024tok, Etheredge:2024amg, Etheredge:2025ahf}.\footnote{For taxonomic constraints on the moduli space metric, see  \cite{Aoufia:2026upcoming}.}
Note that, whereas we have defined the $\alpha$-vectors with respect to the canonical field basis $\phi$, the product rules are basis independent and therefore also hold
with respect to the field $t$.  For example the lightest tower with mass $m_\text{lightest}$ is satisfying the following product rule
\begin{align}
	(\nabla \log m_\text{lightest})^2=\frac{D-2}{(D-d)(d-2)},
\end{align}
where $D$ is the number of spacetime dimensions of the higher theory, and $D\rightarrow \infty$ for the perturbative string limits. This dot product relation results in the commutation relation
\begin{align}
	\left [\log m_\text{lightest},\frac d{dx^0}\log  m_\text{lightest}\right ]=i\frac{D-2}{(D-d)(d-2)}\, .\label{e.lightest}
\end{align}

Another example is the relation between the species scale and the mass of the lightest tower. From the ESC, one has that the lightest tower and the species number $N_s$ (which is related to the species scale $\Lambda_s$ by  $\Lambda_s=N_s^{1/(2-d)} $) satisfy the CRV Tower-Species pattern \cite{Castellano:2023stg,Castellano:2023jjt}
\begin{align}
	\nabla \log m_\text{lightest}\cdot \nabla \log N_s=-1,\label{CRV}
\end{align}
and this translate to commutation relation
\begin{align}
	\left[\log N_s,\frac d{dx^0}\log m_\text{lightest} \right]=-i.\label{fhcom2}
\end{align}
This commutation relation was derived in~\cite{Anchordoqui:2025izb} and implies that $\log N_s$ and $\frac d{dx^0}\log m_\text{lightest} $ form a canonically conjugate pair.

There exist generalizations of equations \eqref{e.lightest} and \eqref{fhcom2}. For example, in a decompactification limit from a $d$-dimensional theory to a $D$-dimensional theory, the $q$-dimensional branes of the theory (that are not necessarily the lightest tower) have discretized dot products with the masses of the lightest towers given by the radion lattice equation of \cite{Etheredge:2025ahf}.
In a decompactification limit, a $(q-1)$-brane has a tension $T_q$ governed by the radion-lattice formula of \cite{Etheredge:2025ahf}
\begin{align}
\begin{aligned}
	(\nabla \log m_\text{lightest})\cdot(\nabla \log T_q)&=\frac{q(D-2)-P(d-2)}{(D-d)(d-2)} \, ,
\end{aligned}
\end{align}
where $P$ is a non-negative integer. 
In species quantum mechanics this relation becomes the following commutator 
\begin{equation}
\left\lbrack \log m_\text{lightest}, {d\over dx^0}\log
  T_\text{subleading} \right\rbrack \!\!=
i\frac{q(D-2)-P(d-2)}{(D-d)(d-2)} \,
 .\label{e.subleading}
\end{equation}
	
For the case where $q=1$, $T_q$ is just a mass. For the case that $T_q$ is actually the leading tower
(i.e. a KK-mode), $P=0$ and \eqref{e.subleading} reproduces \eqref{e.lightest}, but for a particle coming from a wrapped $(P-1)$-brane in the higher dimensional theory, $P$ can be greater than 0. In particular,  for particle towers coming from particle towers in the $D$-dimensional theory, $P=1$, and $m_\text{subleading}$ and $\Lambda_s$ scale at the same rate, and this reproduces \eqref{CRV}. 
As discussed in \cite{Etheredge:2025ahf}, $P$ such that $p\leq P\leq D-2$, $P$ can be interpreted as the spacetime dimension of the brane in the higher-dimensional theory, but $P$ can also take more exotic values, such as when describing KK-modes, KK-monopoles, and exotic branes. Analogous formulas hold for perturbative string limits (see the radion lattice formula of \cite{Etheredge:2025ahf}).

These formulas are for single directions in moduli space. However, one
can generalize the discussion and consider multiple different
directions in a flat slice of moduli space. In particular, given a set
of directions in a common duality frame on a flat slice of moduli
space, one can have multiple radions and dilatons parameterise this
slice, and the leading tower for each radion or dilaton provides a set
of dot-product rules. These towers also satisfy dot product rules with
each other, leading to the lattice formulas discussed in~\cite{Etheredge:2025ahf}, with a generalized dot product rule.

\subsection{Assumptions and limitations}

Many of these patterns typically do not hold in the bulk of moduli space, and also do not hold in special classes of asymptotic limits of moduli spaces. For instance, as was found in \cite{Etheredge:2023odp}, often gradients of logarithms of the lightest particle towers are not constant asymptotically, and ``slide". This happens during running decompactification. In these cases, the taxonomy rules have not been yet determined, and the naive versions above are not correct. It is also quite likely that the CRV pattern does not hold in this context. This is because the CRV pattern, like the other taxonomy rules, follows automatically from the ESC when sliding is not present. But, when sliding is present, the CRV pattern does not automatically hold, and thus likely fails in similar ways in which the other taxonomy rules fail.

Furthermore, except in special cases (involving large amounts of supersymmetry and BPS branes), these taxonomy rules do not hold in the bulk of moduli spaces. For instance, if one defines the species scale as the coefficient of the $R^4$ term in the low energy effective action, then, as one ventures from an asymptotic limit of moduli space towards the bulk of moduli space, it has been conjectured that the lightest tower and the species scale satisfy the inequality
\begin{equation}
  {\nabla \log m_\text{lightest}} \cdot {\nabla \log \Lambda_s} \leq
  \frac{1}{d-2} \, ,
\label{CRVbulk}
\end{equation}
in the interior of moduli space~\cite{Bedroya:2024uva,Basile:2025bql}. Thus, by combining (\ref{fhcom2}) and 
(\ref{CRVbulk}) the
operators within the commutator of (\ref{fhcom2}) may 
have a reduced level of quantum uncertainty in the bulk compared to that at the boundary. 
In addition, in the interior of the moduli space, the product on the l.h.s. of the CRV relation (\ref{CRV})  is in general not anymore constant, which implies that in the interior of moduli space the r.h.s. of the commutator (\ref{fhcom2}) 
is not anymore a c-number but given by a specific operator.

\subsection{Commutation relations and patterns with potential}

\label{sec:2 potential}

In this section we are interested in establishing the commutation
relations between moduli dependent potential $V$ and the species scale $\Lambda_s$, respectively the species number $N_s$. Down this trail, a  pattern similar to
(\ref{CRV}), but now in terms of $V$ has been discussed
in~\cite{Bedroya:2025ris,Bedroya:2025ltj,Bedroya:2025fie} for positive and negative potentials, 
\begin{equation}
  \frac{\nabla V}{V} \cdot \frac{\nabla \Lambda_s}{\Lambda_s} \leq
  \frac{2}{d-2} \,.
\label{ANSS}
\end{equation}
In the following we turn to reexamine this relation.

\subsubsection{AdS vacua and negative potentials}

The asymptotic no-scale-separation (ANSS) condition states that for
any scalar potential with an AdS critical point, there exists an
infinite-distance limit along $\nabla V$ in which the scalar potential
remains negative and (\ref{ANSS}) is satisfied~\cite{Bedroya:2025ltj}.

To examine the ANSS further we first start with 
 the anti-de Sitter (AdS) distance conjecture (ADC) \cite{Lust:2019zwm}, which states
 that for AdS vacua in quantum gravity the limit of a small (negative)
 cosmological constant $\Lambda_\text{AdS}$ is always
accompanied by a light tower of states of mass scale $m_t$. This implies again a relation of the form
\begin{equation}
m_t=|\Lambda_\text{AdS} |^{1/a}\,, \label{lambdatower}
\end{equation}
where the scaling parameter $a$ should be greater or equal than two:
$a\geq2 $. The case $a=2$ is special, it is called the strong ADC,
since then there is no scale separation between the cosmological
constant $\Lambda_\text{AdS}$ and the
tower mass scale $m_t$. Actually, this case of no scale separation is
realised for many AdS flux vacua like $AdS_5\times S^5$ or $AdS_{4(7)}\times S^{7(4)}$. When compared to the situation of a positive potential, we see that the case
of no scale separations corresponds to the saturation of the Higuchi
bound~\cite{Higuchi:1986py}. In addition to the AdS vacua with no scale
separation, potential AdS flux vacua with scale separation and $a>2$,
called DGKT-like vacua \cite{DeWolfe:2005uu}, have been discussed in the literature.

Note that the ADC relates $\Lambda_\text{AdS}$, which is the value of the  minimum of a certain scalar potential $V(\phi)$, to the mass scale of the tower of states. Therefore varying $\Lambda_\text{AdS}$ and considering the limit $\Lambda_\text{AdS}\rightarrow 0$
means that one is comparing a family of AdS vacua, which are characterised for example by different choices of (discrete) flux numbers.
In contrast to that, the ANSS condition is dealing with an off-shell potential $V(\phi)$, which possesses a single vacuum that respects the ADC conjecture for finite scalar field value $\phi_{min}$ and goes to zero for infinite scalar field values,
i.e. $V(\phi)\rightarrow 0_-$ for $\phi\rightarrow \infty$.
We will now assume that also the asymptotic off-shell potential $V(\phi)$ is following the ADC conjecture (\ref{lambdatower}). Combining it with the ANSS condition (\ref{ANSS}) one again derives the following commutator relation
for negative potentials 
\begin{equation}
\left[\log N_s, \frac{d}{dx^0} \log \sqrt{-V}\right] =- {2i\over a}\, .
\label{ansscommutator3}
\end{equation}
We see that the commutator reaches its maximal, canonical form for
$a=2$, which corresponds to the case of no scale separation. In this
case, the relation (\ref{ansscommutator3}) is equivalent to the CRV commutator in (\ref{fhcom2}).

\subsubsection{Positive potentials}

We next discuss the case of positive scalar potentials with $V(\phi)\geq 0$.  We will not assume that there are stationary de Sitter minima with
$V_{min}\equiv\Lambda_{dS}\neq 0$ and $\nabla V=0$. Actually these de Sitter minima would be excluded provided that the de Sitter conjecture~\cite{Obied:2018sgi} is valid~\cite{Agrawal:2018own}. 
So $V(\phi)$ is in general some non-constant potential and approaches  $V(\phi)\sim \exp(-\alpha_V\phi)$ in the infinite distance limit, where the scalar field is asymptotically rolling down towards $\phi\rightarrow\infty$.
These kind of potentials are relevant for cosmology (see e.g.~\cite{Anchordoqui:2021eox,Anchordoqui:2025fgz,Anchordoqui:2025epz,Andriot:2025los,Bedroya:2025fwh,Anchordoqui:2026hys}), where the background is time dependent, and therefore also the potential depends on time, i.e. $\partial_{x^0}V(\phi(x^0))\neq0$.

We start by substituting $\nabla V/V = 2 \nabla \!\sqrt{V}/\sqrt{V}$ into (\ref{ANSS}) we find  
that the ANSS-like relation takes the form
 \begin{equation}
\frac{\nabla \!\sqrt{V}}{\sqrt{V}} \cdot \frac{\nabla \Lambda_s}{\Lambda_s} \leq \frac{1}{d-2} \, .
\label{halfANSS}
\end{equation}
Equivalently, using again
$\Lambda_s = N_s^{1/(2-d)}$, (\ref{halfANSS}) can be rewritten as
\begin{equation}
\frac{\nabla \! \sqrt{V}}{\sqrt{V}} \cdot \frac{\nabla N_s}{N_s} \geq - 1  \, .\label{halfANSS1}
\end{equation}
In the essence of species quantum mechanics~\cite{Anchordoqui:2025izb} we then derive the following ANSS-like commutator
\begin{equation}
-i\left[\log N_s, \frac{d}{dx^0} \log \sqrt{V}\right] = \frac{\nabla
  N_s}{N_s} \cdot \frac{\nabla \! \sqrt{V}}{\sqrt{V}} \geq - 1\, .
\label{ansscommutator}
\end{equation}
 This implies that if the inequality of (\ref{ANSS}) is
saturated, then the
commutator between the two operators in (\ref{ansscommutator}) is constant
and also entails that 
$N_s$ and the time derivative of $\sqrt{V}$ are related to canonically
conjugate operators.

For a positive potential, we can consider the dark dimension relation between the potential and the tower mass scale of the following form:
\begin{equation}
m_t(\phi)=V(\phi)^{1/a}, \quad 2 \leq  a \leq d\,,\label{vtower}
\end{equation}
where the lower limit of the $a$ range comes from the saturation of the Higuchi bound~\cite{Higuchi:1986py},
and the upper limit comes from the Casimir energy in $d$
dimensions~\cite{Montero:2022prj}.  Using the asymptotic CRV pattern relation (\ref{fhcom2}),
the ANSS-like commutator (\ref{ansscommutator}) can be written as
\begin{equation}
\left[\log N_s, \frac{d}{dx^0} \log \sqrt{V}\right] =- {2i\over a}\, .
\label{ansscommutator1}
\end{equation}
So this commutator 
 reaches its maximum possible extent through saturation of
the Higuchi bound, where $m_t^2=V$. 
For larger values of $a$, the commutator gets suppressed, meaning that there is a reduced amount of quantum uncertainty.

Recall that a non-vanishing commutator requires that the potential is not a constant, but varies with respect $\phi$ inducing in this way a $x^0$-dependence. 


In standard 4D cosmology the rate
at which the universe is stretching apart is measured by the Hubble
parameter,
\begin{equation}
  H \sim \sqrt{V} \, .
\label{HV}
\end{equation}
Now, the derivative of the logarithm of $H$
(normalized by $H$) typically relates to the universe's
acceleration/deceleration parameter
\begin{equation}
  q = -1 - \frac{\dot H}{H^2} \,,
\label{qdef}
\end{equation}
so
\begin{equation}
  \frac{d(\log H)}{dx^0} = \frac{\dot H}{H} = - H(1 +q)  \, .
  \label{ansscommutatorf}
\end{equation}
Note that if $q=-1$ or equivalently $d (\log H)/ dx^0 =0$, the Hubble parameter
is constant leading to exponential expansion, which occurs in a vacuum
dominated dS space. This would correspond to a constant $V$, in which
case  $d (\log \sqrt{V})/ dx^0$ would commute with
$\log N_s$, collapsing into a classical solution.
Finally, the ANSS-like commutator (\ref{ansscommutator}) can be rewritten as
\begin{equation}
  \left[\log N_s, H(1+q)\right] =
     {2i\over a}
              \, .
\label{ansscommutator2}
\end{equation}
The operators within the commutator are canonically conjugate if $a=2$.  The case of $q=-1$ corresponds to a constant vacuum energy
without field dependence, see the Appendix A for details.

\section{Moduli space wave functions}

\label{sec:3}

In this section, we examine wave-functions on moduli spaces, both with and without potentials. In particular, we find that the quantum mechanics and geometry of moduli space result in excited states that have moduli localised around regions that are not the classical minima of the potentials.

\subsection{The free case without potential}

We begin by analyzing the free Schr\"odinger equation, which
characterizes a particle unaffected by external forces or potential
energy, defined by the condition $V(t) = 0$.




\subsubsection{One-dimensional moduli space}

As explained in~\cite{Anchordoqui:2025izb}, for stationary states
(i.e. states with definite energy $E$), we adopt the time-independent
version of the free Schr\"odinger equation that takes the form
\begin{equation}
{\cal H} \psi_{E}=E \psi_{E}\, ,\label{schrodinger}
\end{equation}
where the wave functions $\psi_{E}$ are the eigenfunctions  of the
Hamiltonian ${\cal H}$. Since we are assuming that the particle moves
without external influence, the Hamiltonian is just given by the  negative of the Laplacian as:
\begin{equation}
{\cal H}=-{\nabla^2\over 2}=-{\partial^2_\phi\over 2}\, .
\end{equation}
Its eigenfunctions are given by the following exponential function:
\begin{align}
	\psi_{\alpha}(\phi)=c\exp(\pm i\alpha \phi)\, ,
\end{align}
where $\alpha$ plays the role of the momentum eigenvalue, $c$ is a
normalisation constant, and the corresponding energy eigenvalue is given by 
\begin{equation}
E={\alpha^2\over 2} \, .
\end{equation}

In terms of $t$, 
the moduli space wave function is proportional to 
\begin{equation}
\psi_\alpha(t) \sim \exp\left[\pm i\sqrt k\alpha \log t\right]=t^{\pm i\sqrt k\alpha}\,. 
\end{equation}
This plane wave function is non-normalizable and non-square integrable over the entire moduli space. 

We can  also compare the wave function with the species scale, which is asymptotically a decaying function of the form,
\begin{equation}
\Lambda_s(t)\sim t^{-{\sqrt{k}\alpha_s}}=e^{-\alpha_s\phi}\, ,
\end{equation}
or with the lightest tower,
\begin{equation}
m(t)_\text{lightest}\sim t^{-{\sqrt{k}\alpha_\text{lightest}}}=e^{-\alpha_\text{lightest}\phi}\, ,
\end{equation}
where the decay parameters $ \alpha_s$ and $\alpha_\text{lightest}$
obey the taxonomy relation $ \alpha_s \ \alpha_\text{lightest}= 1/
  (d-2)$.
So we see that the species scale and the species wave function can be obtained from each other
by replacing the momentum eigenvalue $\alpha$  of $\psi_\alpha$ by an
imaginary momentum, namely $\alpha \rightarrow\pm i\alpha_s$. In other words, {\it the species scale takes the form of a particular
  Wick rotation in momentum space of its wave function.}

\vskip0.3cm\noindent
{\underline {\bf Ground state:}}

\vskip0.3cm

The ground state of the quantum system has zero momentum,
i.e. $\alpha=0$. Hence, the ground state wave function is a constant. The momentum uncertainty is zero, $\Delta\alpha=0$,
and hence the position uncertainty is infinite, $\Delta
t=\infty$. Such  state of perfectly defined (or {\it sharp}) momentum also implies that the uncertainty of the species scale is infinite and one cannot determine where its wave function is located. The vanishing of the momentum
uncertainty ensures that the position in moduli space is undetermined
and $t$ is a flat direction. This is in agreement with what one knows
from  conformal field theories and  string theory, where the field $t$
(or respectively $\phi$)
is massless and corresponds to a marginal
operator hence to a flat direction. This is also reflected by the ground state wave function of moduli space  quantum mechanics.

\vskip0.3cm\noindent
{\underline {\bf Excited states:}}

\vskip0.3cm

The excited states are plane waves $\psi_\alpha(t) \sim t^{\pm i\sqrt k\alpha}$ with non-vanishing momentum $\alpha$.
In  case there is a duality symmetry of the form $t\rightarrow
1/t$, the plane wave contains two terms,
\begin{equation}
\psi_\alpha(t) \sim at^{i\sqrt k\alpha}+bt^{-i\sqrt k\alpha}\quad (|a|=|b|)\, ,
\end{equation}
and travels towards the self-dual point $t=1$ and bounce back at $t=1$ to further travel towards $t\rightarrow\infty$.
For excited states with non-vanishing momentum $\alpha$, one obtains  
 normalizable wave functions by  constructing  wave packets with a finite extent in moduli space (e.g. a Gaussian wave packet), which are superpositions of different momentum (or wave number $\alpha$) eigenstates,
\begin{eqnarray}
\psi(t) & \sim & \frac{1}{\sqrt{2\pi}} \int_{-\infty}^{+\infty}
                   \varphi (\alpha) \ e^{i\alpha \phi(t)} \ d\alpha \nonumber \\
& \sim & e^{-\frac{[\phi(t)-\phi(t_0)]^2}{2 (\Delta t)^2}} \ e^{i\alpha_0 \phi(t)} \,,
\end{eqnarray}
where $\Delta t$ determines the width of the packet around $t_0$.
In this case, there is an uncertainty relation
\begin{equation}
\Delta (\log t)\Delta \alpha\geq 1\, ,
\end{equation}
where $\Delta\alpha$ is the momentum uncertainty around $\alpha_0$. As usual, the uncertainty relation is saturated for Gaussian wave packets.




\subsubsection{Higher-dimensional moduli space with finite volume}

As discussed in \cite{Hamada:2021yxy,Delgado:2024skw}, the moduli space in quantum gravity must be compactifiable and its volume must be finite or at least grow no faster than that of Euclidean space. 
More concretely, for a particular cut-off $\Lambda$ of the EFT, one has to restrict the moduli space to a subset ${\cal M}_\Lambda$ for which the cut-off is smaller than the species case, i.e. $\Lambda<\Lambda_s(\phi)$ 
\cite{Hamada:2021yxy,Delgado:2024skw}. Then for any EFT with finite cut-off $\Lambda$ the truncated moduli space has finite volume:
\begin{equation}
V({\cal M}_\Lambda)<\infty\, .
\end{equation}

The one-dimensional moduli space, discussed in the previous section, is flat and has infinite volume. It just corresponds to the limiting case of Euclidean space. As we have discussed the corresponding quantum wave function is
given by a plane wave and hence non-normalizable and non-square integrable. We now proceed with a two-dimensional moduli space, which has finite volume. 
In this case we will require that  wave functions that live in the two-dimensional moduli space must be
normalisable and square integrable.

\begin{figure}
\begin{center}
\includegraphics[width = 80mm]{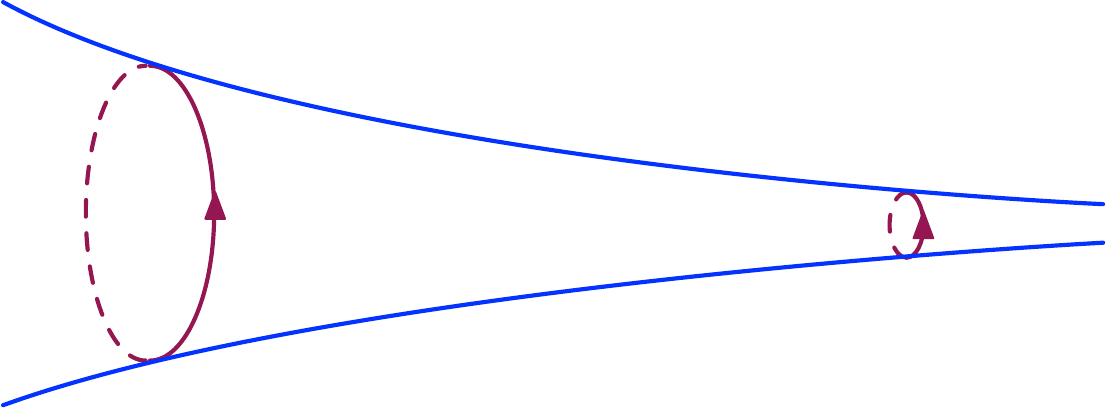}
\end{center}
\caption{A saxionic limit of moduli space. The circular periodicity (highlighted in purple), has radius controlled by the axion decay constant, and this becomes exponentially small as the saxion increases. This exponential decay squeezes the normalizable wavefunctions into the bulk.}
\label{f.saxion}
\end{figure}

Specifically consider a simple two-dimensional moduli space locally of the form ${\mathbb R} \times S^1$, where the metric is
\begin{equation}
{\cal M}:\qquad ds^2 = d\phi^2 +k e^{-{2\gamma\over \sqrt k}|\phi |}d\theta^2\, ,\label{2dimmetric}
\end{equation}
where $\phi\in{\mathbb R}$  is the non-compact coordinate, $\theta = \theta +2\pi$ is the periodic coordinate, and $\gamma$ is a
positive constant that controls the exponential rate at which the circle's radius shrinks (see Fig. \ref{f.saxion}).
The moduli space is symmetric under the duality symmetry $\phi\rightarrow-\phi$.
As discussed in \cite{Delgado:2024skw}, the corresponding truncated moduli space ${\cal M}_\Lambda$ has indeed
finite volume. 

Performing the already introduced field redefinition of the form
\begin{equation}
\phi={\sqrt k}\log t\, ,
\end{equation}
we introduce a complex coordinate
\begin{equation}
\tau=\theta +i t=\theta +ie^{{1\over \sqrt k}\phi}\, .
\end{equation}
Then the moduli space metric becomes
\begin{equation}
{\cal M}:\qquad ds^2 = k\biggl({dt^2\over t^2} + {d\theta^2\over t^{2\gamma}}\biggr)\, .
\end{equation}
For $\gamma=1$, the moduli space is the well-known hyperbolic plane ${\cal M}={\mathbb H}^+$.
 In this case the metric takes its quite familiar form
\begin{equation}
{\cal M}:\qquad ds^2 = k{d\tau d\bar \tau\over {\Im \tau}^2} \, .
\end{equation}
This is just the K\"ahler metric of the space ${SL(2,{\mathbb R})\over U(1)}$. Due to the duality symmetries $\tau\rightarrow {a\tau+b\over c\tau+d}$ one has to further mod out by the action of the 
$SL(2,{\mathbb Z})$ discrete duality group. Then the moduli space has the well-known form of the fundamental domain of $SL(2,{\mathbb Z})$.
We will study the restrictions from modular invariance in more detail in the next subsection.

\vskip0.3cm

We will now investigate the solutions of the two-dimensional Schr\"odinger equation 
\begin{equation}
-{\nabla^2\over 2} \psi_{E}(\phi,\theta)=E \psi_{E}(\phi,\theta)\,.
\end{equation}
The negative Laplacian of the metric (\ref{2dimmetric}) has the form
\begin{equation}
-\nabla^2 =-\partial^2_\phi+{\gamma\over\sqrt k}\partial_\phi-{1\over k}e^{{2\gamma\over\sqrt k}|\phi |}\partial^2_\theta\, .
\end{equation}
We require the wave functions
$\psi_{E}(\phi,\theta) $ to be single-valued under $\theta\rightarrow\theta+2\pi$. This is accomplished with wave functions of the form
\begin{equation}
\psi_{E}(\phi,\theta)=\sum_{n\in{\mathbb Z}}c_ne^{in\theta}\psi_n(\phi)\, ,\label{ansatz}
\end{equation}
where $c_n$ are constants. Schr\"odinger's equation implies that each $\psi_n$ satisfies
\begin{equation}
-{1\over2}\psi_n''+{\gamma\over2\sqrt k}\psi_n'+{n^2\over 2k}e^{{2\gamma\over\sqrt k}|\phi |}\psi_n=E\psi_n\, ,
\end{equation}
where primes denote $\phi$-derivatives.
It is convenient to define
\begin{equation}
\chi_n(\phi)=e^{-{\gamma\over2 \sqrt k}|\phi |}\psi_n(\phi)\, ,\label{ansatz1}
\end{equation}
and the Schr\"odinger equation simplifies to
\begin{equation}
-{1\over 2}\chi_n''+{n^2\over 2k}e^{{2\gamma \over\sqrt k}| \phi |}\chi_n={\alpha^2\over 2}\chi_n\, ,
\end{equation}
with momenta $\alpha$ and energy
\begin{equation}
E={\gamma^2\over 8k}+{\alpha^2\over 2}\, .
\end{equation}
This is just the 
familiar 
Schr\"odinger equation for a particle on a line, and with an potential of the form
\begin{equation}
V_{\rm geo}(\phi)={n^2\over 2k}e^{{2\gamma\over\sqrt k}| \phi |}\, .\label{effpot}
\end{equation}
This ``geometric" potential originates from the non-flat geometry of the moduli space and acts as attracting the quantum particle towards the bulk of the moduli space.

\vskip0.3cm\noindent
We now describe the moduli space wave functions and the spectrum of the system.

\vskip0.3cm\noindent
{\underline {\bf Ground state:}}

\vskip0.3cm
The ground state has  $n=0$ and hence the effective potential is absent.
Then the wave function does not depend on $\theta$, and as in the one-dimensional case the wave functions are planes waves proportional to $\psi_0\sim t^{i\alpha}$.
Furthermore, the ground state has momentum $\alpha_0=0$.
This is analogous to the one-dimensional moduli space,  $\phi$ and $\theta$ are undetermined for $\alpha_0=0$ and correspond to the flat directions of the moduli space. 
Actually the ground state is simply given, in addition to the wave functions in eq.(\ref{ansatz1}), by the constant wave function $\psi_0={\rm const}$, which is also modular invariant (see next section) and has zero energy:
\begin{equation}
E_0=0\, .
\end{equation}
The integral of $\psi_0^2$ over the moduli space ${\cal M}$ is just the volume of ${\cal M}$, so for the truncated moduli space ${\cal M}_\Lambda$ of finite volume the ground state is normalizable.

\vskip0.3cm\noindent
{\underline {\bf Excited states:}}

\vskip0.3cm

Next, consider the excited states. These are the eigenfunctions $\psi_{E}$ of the Laplacian with $\alpha,n\neq0$. Then there is also the effect from the non-vanishing geometric  potential as given in Eq.~(\ref{effpot}).
This potential has a minimum at the self-dual point $\phi=0$. So for the excited states the modulus $\phi$ will be stabilised near this point by quantum effects - see also the further discussion about the modular invariant wave function.
Furthermore, the excited states have positive energy.

The excited wave functions $\chi_n(\phi)$ will have non-vanishing support in the entire two-dimensional moduli space. Hence they must be  square integrable.
This requires that 
\begin{equation}
\int d\phi|\chi_n(\phi)|^2<\infty\, .
\end{equation}
Concretely, the  functions $\chi_n$ are given in terms of modified Bessel functions $K_\nu$. For large $t\rightarrow\infty$    they decay exponentially with respect to $t$ and hence double exponentially with respect to 
the limit  $\phi\rightarrow \infty$ and they have the  form
\begin{equation}
\chi_n(\phi)=C K_\nu\biggl({2\pi\sqrt k|n|\over \gamma}e^{{2\gamma\over \sqrt k}\phi}\biggr)\quad{\rm for}\quad \phi>0\, ,\label{specieswavefunction}
\end{equation}
where $\nu=\sqrt{-\alpha^2}=i\alpha$.  Note that the momenta $\alpha$ can only take certain discrete values (see below). The energies of the excited states are  
\begin{equation}
E={\gamma^2\over 8k}+{\alpha^2\over 2}\, .\label{spectrum}
\end{equation}
Note that the energy spectrum is strictly positive.

\vskip0.3cm

In the other, dual  branch $\phi<0$, the wave functions 
decay exponentially for $\phi\rightarrow -\infty$ for $\phi<0$ and are given as
\begin{equation}
\chi_n(\phi)=C K_\nu\biggl({2\pi\sqrt k|n|\over \gamma}e^{-{2\gamma\over \sqrt k}\phi}\biggr)\quad{\rm for}\quad \phi<0\, ,
\end{equation}

\subsubsection{Modular invariant wave functions}


We now consider the case  $\gamma=1$ and $k=1/2$ in more detail. Here we will require that the wave functions are modular invariant.
In general, the eigenfunctions of the $SL(2,{\mathbb Z})$ invariant Laplacian $\nabla^2$ are given in terms of the non-holomorphic Eisenstein series $E_s(\tau,\bar \tau)$, which can be defined as 
\begin{equation}
E_s(\tau, \bar\tau)=\sum_{p,q\in{\mathbb Z}}^{'}{t^s\over\pi^s|p+q\tau|^{2s}}\, ,
\end{equation}
where $t=\Im\tau$ and the prime on the sum denotes that $p=q=0$ is omitted from the sum. In string theory this sum can be viewed as the sum over the tensions of $(p,q)$-strings, raised to the power $-2s$.

The non-holomorphic Eisenstein series  are special examples of a more general class of non-holomorphic, modular invariant objects known as Maa{\ss} forms,  denoted by ${\cal N}(s)$ 
(see for example \cite{DHoker:2022dxx,Aoufia:2025ppe,Aoufia:2026upcoming}).
Each Maa{\ss} form $f_s(\tau,\bar\tau)$  satisfies a Laplace eigenvalue equation 
\begin{equation}
{1\over 2}\nabla^2f_s(\tau ,\bar\tau)=\lambda f_s(\tau,\bar\tau)\, ,\qquad \nabla^2=2t^2(\partial_{\theta}^2+\partial_{t}^2)\, ,
\end{equation}
where the corresponding eigenvalues can be written as 
\begin{equation}
\lambda=s(s-1)\, ,
\end{equation}
with $s$ being a complex number.
A general element of ${\cal N}(s)$ has a Fourier expansion of the form
\begin{equation}
f_s(\tau,\bar\tau)=at^s+bt^{1-s}+\sum_{n\neq0}c_n\sqrt{t}e^{i\pi n\theta}K_{s-1/2}\bigl(2\pi|n|t\bigr)\, .\label{maass}
\end{equation}
This expansion matches the Eisenstein series upon the following identification:
\begin{eqnarray}
&{~}&a={2\zeta(2s)\over \pi^s}\, ,
\quad 
b={2\Gamma(s-{1\over 2})\pi^{{1\over 2}-s}\zeta(2s-1) \over\Gamma(s)}\nonumber\\
&{~}&c_n={4|n|^{{1\over 2}-s}\sigma_{2s-1}(|n|) \over\Gamma(s)}
\end{eqnarray}
The polynomial term,  $at^s+bt^{1-s}$, in the above Fourier expansion is called Laurent polynomial of the Maa{\ss} form.
It prevents the 
polynomial piece to be square integrable.
One can subtract the Laurent polynomial from the Maa{\ss} form, and then at least for certain values of $s$ (see \cite{DHoker:2022dxx} for details) one obtains a square integrable cusp form ${\cal S}(s)$.

Now let us determine, by which Maa{\ss} form resp. by which cusp form the moduli space wave function $\psi_{E}$ is given.
For that we will compare the relevant energy eigenvalues. In moduli space quantum mechanics 
 we are looking for real and negative eigenvalues of the Laplacian, i.e. we have the following identification:
 \begin{equation}
 E=-\lambda\quad\Longleftrightarrow\quad {1\over 4}+{\alpha^2\over 2}=-s(s-1)\, .
 \end{equation}
This equation is solved by  {\sl complex} values for $s$, namely by setting
\begin{equation}
s={1\over 2}+i{\beta}\, ,\quad \beta={\alpha\over\sqrt2}\, .
\end{equation}
So we 
arrive at the following identification for  moduli space function:
\begin{eqnarray}
\psi_{\beta}(\tau,\bar\tau)=E_{{1\over 2}+i\beta}(\tau,\bar\tau)=
\sqrt{t}\biggl(at^{i\beta}+bt^{-i\beta}+\sum_{n\neq0}c_ne^{i\pi n\theta}K_{i\beta}\bigl(2\pi|n|t\bigr)\biggr)\, .\label{wavefunctionmod}
\end{eqnarray}

One can easily check that $E_{{1\over 2}+i\beta}(\tau,\bar\tau)$
matches the previously derived expressions of the  wave function, as
given in Eqs.~(\ref{ansatz}), (\ref{ansatz1}), and (\ref{specieswavefunction}).
Note that with $\Re s=1/2$ there is in all terms of the expansion (\ref{wavefunctionmod}) a common factor $\sqrt t$, which precisely agrees with $e^{\phi/\sqrt2}$ factor in Eq.~(\ref{ansatz1}).

\vskip0.3cm
The plane wave functions of the system are given in terms of the Laurent polynomial,
 \begin{equation}
 \psi_{\beta}^{\rm wave}(t)=\sqrt{t}(at^{i\beta}+bt^{-i\beta})\, .
 \end{equation}
  This is an oscillatory wave with momentum $\alpha$, which hits a wall at $|\tau|=1$, where it is reflected and is asymptotically given as 
$\sqrt{t}(t)^{-i\beta}$ (see also \cite{DHoker:2022dxx}).

In addition to the wave solutions, the excited spectrum includes bound states that
correspond to the square integrable cusp forms 
\begin{equation}
\psi_\beta^{\rm bound~state}(t)={\cal S}(1/2+i\beta)=\sqrt{t}\sum_{n\neq0}c_ne^{i\pi n\theta}K_{i\beta}\bigl(2\pi|n|t\bigr)\, .
\end{equation}
The bound state wave functions decrease exponentially towards $t\rightarrow\infty$. 
Actually, the bound states with positive energy correspond to particles that are caught inside the moduli space ${\mathbb H}^+$ that becomes thinner and thinner towards $t\rightarrow\infty$.
This localization of the wave function is precisely the effect of the
geometric potential in Eq.~(\ref{effpot}).\footnote{Following~\cite{Cooper:1994eh}, we expect the supersymmetric spectrum to also exhibit bound states, which undergo a shift in energy levels upon the inclusion of fermions.}

There are infinitely many bound states with an energy spectrum (\ref{spectrum}) that goes up to infinity.
One can show that the momentum eigenvalues of the bound states $\alpha_i$ are discrete, but they can be calculated only approximately as \cite{DHoker:2022dxx}
\begin{equation}
\beta_1\sim 9.53\, ,\quad \beta_2\sim 12.17\, ,\quad\beta_3\sim 13.78\, \dots
\end{equation}

We next consider the vacuum expectation values for $t^n$, i.e. for the n-th. positive power of the position operator $t$ in moduli space. This expectation value is given by the following expression
\begin{equation}
\langle t^n\rangle_\beta=\langle \psi_\beta|t^n|\psi_\beta\rangle=\int_{\cal F}t^n|E_{{1\over 2}+i\beta}|^2{3\over \pi}{d\theta d t\over t^2}\, ,
\end{equation}
where ${\cal F}$ is the fundamental domain of the moduli space and  $d \mu={3\over \pi}{d\theta d t\over t^2}$ is its invariant measure.
For the wave-like states we have that $|E_{{1\over 2}+i\beta}|^2\sim t$, and hence the integral diverges for any $n$, meaning that for the waves the vacuum expectation value $\langle t^n\rangle_\beta$ is undermined.

However for the bound states the above integral is finite, since the corresponding wave functions are exponentially suppressed for large $t$. E.g. for $n=1$, the vacuum expectation value of the position operator for the first eigenvalue 
$\beta_1\sim 9.53$ is close to the center of moduli space and can
be numerically computed to be \
\begin{equation}
\langle t\rangle_{\beta_1}\sim 1.2 - 1.4\, .
\end{equation} 
For the higher eigenvalues, the $\langle t\rangle_{\beta_i}$ can penetrate deeper into the moduli space, where however the growth of the expectation values is
rather slow.
In physical terms it means that for all bound states the modulus $t$ is stabilised at increasing but finite values inside the moduli space.
Finally we can also consider the vacuum expectation value for $\langle \phi\rangle_{\beta_i}=\langle \log t\rangle_{\beta_i}$. The corresponding  integral is finite and numerically gives a value of order
\begin{equation}
\langle \log t\rangle_{\beta_1}\sim 0.5 - 0.6\, .
\end{equation}

\subsubsection{Comparison with the species scale}


\vskip0.3cm\noindent
{\underline {\bf Type IIB in ten dimensions:}}

\vskip0.3cm

Let us contrast the moduli space wave function $\psi(\tau,\bar\tau)=E_{{1\over 2}+i\beta}(\tau,\bar\tau)$ to the moduli dependent  species scale $\Lambda_s(\tau,\bar \tau)$.
For type IIB in ten space-time dimensions and axion-dilaton field $\tau=\theta+ie^{{\sqrt 2}\phi}=\theta+{i\over g_s}$ the eigenfunction of the {\sl positive} Laplacian
with the {\sl  real } value $s=3/2$ is given by the non-holomorphic Eisenstein function $E_{3/2}(\tau)$ \cite{Aoufia:2025ppe}:
\begin{equation}
\nabla^2E_{3/2}(\tau,\bar\tau)={3\over 4}E_{3/2}(\tau,\bar\tau)\, .
\end{equation}
This eigenfunction $E_{3/2}(\tau,\bar\tau)$ is  the Wilson coefficient of the $R^4$-operator in the 10-dimensional IIB effective action~\cite{vandeHeisteeg:2023dlw} (see also Eq.~(\ref{eft}) with $n=8,d=10$). 
It increases
exponentially for $\phi\rightarrow \pm\infty$ and
   it reaches a minimum at $\phi=0$.
 Actually one can show that $E_{3/2}(\tau,\bar\tau))$  is not square integrable.
Moreover $ E_{3/2}(\tau,\bar\tau))$ can be simply obtained from the wave function by replacing the momenta $\beta$ by an imaginary momentum, i.e.  $\beta\rightarrow -i$.

The corresponding  species
scale in ten dimensions   is given as~\cite{vandeHeisteeg:2023dlw}:\begin{equation}
\Lambda_{s}(\tau,\bar\tau)=E_{3/2}(\tau,\bar\tau)^{-1/6}\, .
\end{equation}
 For $t\rightarrow\infty$
  \begin{equation}
  E_{3/2}(\tau,\bar\tau)\rightarrow t^{3/2}\, ,
 \end{equation}
and hence the species scale decays as 
 \begin{equation}
 \Lambda_s(t)\rightarrow t^{-1/4}\, .
 \end{equation}
 This indeed  corresponds to the perturbative string limit in ten dimensions with $k=1/2$:
  \begin{equation}
 \Lambda_s(t)\rightarrow e^{-{1\over \sqrt 8}\phi}\, ,
 \end{equation}
 and hence we confirm that $\alpha_s=\sqrt{1\over d-2}=\sqrt{1\over 8}$.

 For $t\rightarrow\infty$, the wave function goes like 
 \begin{equation}
 \psi(t)\rightarrow  \sqrt tt^{i\beta}\, .\label{waveasym}
 \end{equation}
So in the asymptotic regime, the species scale $\Lambda_s$ is obtained from the wave function $\psi_\beta$ by replacing the real momentum $\beta$ by following imaginary number:
 \begin{equation}
 \beta\, \rightarrow\, {3i\over 4}\, .\label{replace}
 \end{equation}
 
 \vskip0.3cm\noindent
{\underline {\bf Type IIA in ten dimensions:}}

\vskip0.3cm

We can briefly also consider  type IIA in ten dimensions, where the coefficient of the $R^4$-operator is again given by a form $f_{3/2}(\phi)$,  however now with respect to the group $\Gamma=SL(1)$. The corresponding Maa{\ss} form just consists of the 
Laurent polynomial, namely  
\begin{equation}f_{3/2}(t)={3\zeta(3)\over \pi^2}t^{3/2}
+t^{-1/2}\, . 
\end{equation}
The corresponding species scale \cite{vandeHeisteeg:2023dlw} is again $\Lambda_s=(f_{3/2})^{-1/6}$.
In the weak coupling limit $t=e^{\sqrt2\phi}\rightarrow\infty$ we again recover the perturbative string limit with  $\Lambda_s(t)\rightarrow t^{-1/4}$.
On the other hand,
the strong coupling  limit with $t\rightarrow0$
leads to a species scale that goes like 
\begin{equation}
 \Lambda_s(t)\rightarrow t^{1/12}\, .
 \end{equation}
 Here we can define a dual scalar field $\phi_D=-{1\over \sqrt 2}\log t$ and then we obtain that 
 \begin{equation}
 \Lambda_s(t)\rightarrow e^{-{1\over \sqrt {72}}\phi_D}\, .
 \end{equation}
This corresponds to  $\alpha_s=\sqrt{p\over  (D-2)(d-2)}$ with $p=1$, $D=11$ and $d=10$ (see Eq.~(\ref{decomp})) and correctly describes the 11-dimensional M-theory limit of type IIA.

On the other hand, the moduli space wave function of the 1-dimensional IIA moduli space is just a wave of the form
 \begin{equation}
 \psi_{\beta}^{\rm wave}(t)=\sqrt t(at^{i\beta}+bt^{-i\beta})\, .
 \end{equation}
There are no bound states in this case.

\vskip0.3cm\noindent
{\underline {\bf Decompactification from four to six dimensions:}}

\vskip0.3cm

Next we consider decompactification from four to six dimensions, which corresponds to the case $k=1$. Now the complex parameter $\tau$ comprises the volume modulus of a two-torus olus an internal $b$-field.
The moduli space wave function is again given by $\psi=E_{{1\over
    2}+i\beta}$
and contains bouncing wave solutions as well as bound states.
Now the species scale is provided  by the Wilson coefficient of the $R^2$-operator in the 4-dimensional effective effective action (see also Eq.~(\ref{eft}) with $n=4,d=4$). 
This is determined  in terms of the genus-one free energy $F_1$ and it is given by the non-holomorphic Eisenstein function $E_{1}(\tau,\bar\tau)$~\cite{vandeHeisteeg:2022btw,Cribiori:2023sch,vandeHeisteeg:2023dlw}:
\begin{equation}
F_1\simeq E_{1}(\tau,\bar\tau)=
\sum_{p,q\in{\mathbb Z}}^{'}{t\over\pi|p+q\tau|^{2}}\, .
\end{equation}
This form is harmonic, i.e. it is a solution of the Laplace equation with eigenvalue zero.
It is simply derived from the wave function by replacing $\beta\rightarrow -i/2$.
After suitable regularisation $F_1$ can be expressed as 
 \begin{equation}
F_1=-\log\lbrack t|\eta(\tau)|^{4} \rbrack\, ,
\end{equation}
where $\eta(\tau)$ is the Dedekind $\eta$-function. The species scale in $d=4$ is then given as 
\begin{equation}
\Lambda_s=(F_1)^{-1/2}={1\over \sqrt {-\log\lbrack t|\eta(\tau)|^{4} \rbrack}}\,.
\end{equation}
Again we 
check the asymptotic limit for $t\rightarrow\infty$, where one has
\begin{equation}
\Lambda_s(t)\rightarrow t^{-1/2}\, .
\end{equation}
We can replace $t$ by the canonically normalised scalar field $\phi=\log t$, and then we get that 
\begin{equation}
\Lambda_s(t)\rightarrow e^{-\phi/2}\, ,
\end{equation}
from where we read off that $\alpha_s=1/2$. This value indeed corresponds to the correct scaling for the decompactification limit from four to six dimensions with $\alpha_s=\sqrt{p\over  (D-2)(d-2)}$, using
$D=6$, $d=4$ and $p=2$, see eq.~(\ref{decomp}).

\subsection{Non-vanishing potentials}

Now let is consider the case of a non-vanishing potential. For simplicity we consider an exponential potential of the form
\begin{equation}\label{potentialv}
V(\phi)=V_0\exp (-\alpha_V\phi)\, ,
\end{equation}
which is asymptotically of the required form and where $V_0$ and $\alpha_V$ are positive constants.
In order to make the potential duality invariant under $\phi\rightarrow-\phi$, we can add a term like $V_0\exp (\alpha_V\phi)$ and  the potential becomes  \cite{Anchordoqui:2021eox,Anchordoqui:2025fgz}
\begin{equation}
\label{Vinit}
    V =V_0 ~\text{sech}\left(\alpha_V\phi\right) \, .
   \end{equation}
For the two-dimensional moduli space, one can also consider fully $SL(2,{\mathbb Z}) $ invariant potentials of the form \cite{Anchordoqui:2025epz}
\begin{equation}
\label{Vguess}
V(\tau,\bar\tau)  \simeq -\frac{ V_0 }{\log\lbrack t|\eta(\tau)|^{4} \rbrack} \, .
\end{equation}
All these potentials have in common that they have an exponential  
runaway behaviour for large $\phi\rightarrow\pm\infty$.
Therefore for simplicity we will consider in the following the potential (\ref{potentialv}), which is valid in the branch $\phi\geq0$ and in particular in the asymptotic regime $\phi\rightarrow\infty$.
In fact, in the asymptotic regime we can use the taxonomy in order to relate the potential to the asymptotic behaviour of the species scale. Concretely assume that the species number behaves for large $\phi$ as
\begin{equation}
N_s\sim \exp (\alpha_{N_s}\phi)\, ,
\end{equation}
where the constant $\alpha_{N_s}$ depends on the specific tower that becomes massless.
Now using the ANNS relation (\ref{halfANSS1}), we derive that
\begin{equation}
\alpha_V\geq{2\over \alpha_{N_s}}\, .
\end{equation}

The Schr\"odinger equation including the potential is given as 
\begin{equation}
\biggl(-{\nabla^2\over 2} +V_0\exp (-\alpha_V\phi)\biggr)\psi(\phi,\theta)=E \psi(\phi,\theta)\,.
\end{equation}
Going trough the same steps as in the previous section we arrive at the following differential equation for $\chi_n(\phi)$ (for simplicity we have set $k=1$):
\begin{equation}
-{\chi_n''\over 2}+\bigl(n^2e^{2\gamma| \phi |}+V_0\exp (-\alpha_V\phi)\bigr)\chi_n={\alpha^2\over 2}\chi_n\, .
\end{equation}
 Now the effective potential is the sum of the "geometric´´potential and the real potential:
\begin{equation}
V_{\rm eff}(\phi)=n^2e^{2\gamma| \phi |}+V_0\exp (-\alpha_V\phi)\, .\label{effpot1}
\end{equation}

\begin{figure}
\begin{center}
\includegraphics[width = 80mm]{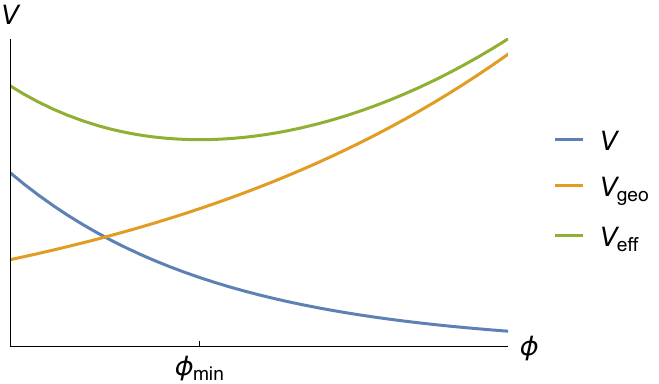}
\end{center}
\caption{The potential $V$, together with the geometric contribution $V_\text{geo}$ together produce an effective potential $V_\text{eff}=V+V_\text{geo}$ with a minimum at a finite region of moduli space.}
\label{f.Veff}
\end{figure}

This is depicted in Fig. \ref{f.Veff}.

\vskip0.3cm\noindent
{\underline {\bf Ground state:}}

\vskip0.3cm

Classically, this potential will not lead to a stable vacuum, but it drags the field $\phi$ to infinity, where the species scale is zero.
The same will happen in the quantum mechanical description. Concretely the ground state wave function has $n=0$, and only the second term in the effective potential (\ref{effpot1})
is present.
The potential is again of the Liouville type, and $V(\phi)$ exponentially decreases for $\phi\rightarrow\infty$.
The ground state wave function is given in terms  a modified  Bessel functions $K_0$ of the form
\begin{equation}
\psi_{0}(\phi)=c ~e^{{\gamma\over 2}\phi}K_0\biggl({1\over \alpha_V}e^{-\alpha_V\phi}\biggr)\, .
\end{equation}
Asympotically for $\phi\rightarrow\infty$ the wave functions become  plane waves with continuous momenta, where the ground state of the system has zero momentum, just like for the free case.

\vskip0.3cm\noindent
{\underline {\bf Excited states:}}

\vskip0.3cm

Now we will show that quantum effects will provide a stable minimum in the $\phi$-direction for all excited states.
For the excited states, $n$ is non-zero and the effective potential (\ref{effpot1})
is now minimized at 
\begin{equation}
\phi_{{\rm min},n}={1\over 2\gamma+\alpha_V}\log\biggl({\alpha_VV_0\over 2\gamma n^2}\biggr)\, .
\end{equation}
This expression is valid for $\phi\geq0$, i.e. for ${\alpha_VV_0\over 2\gamma n^2}\geq1$. Actually, for the higher excited states with larger $n$, the minima are more and more located in the bulk, whereas small $n$ leads to minima
closer to the boundary of field space. All minima have  positive energy eigenvalues.

When one removes the quantum mechanics ($\hbar\rightarrow 0$), or removes the curvature of moduli space
($\gamma\rightarrow0$), the particle approaches infinity and there is no different effective potential. 
Thus, from
the nontrivial geometry and the quantum mechanical effects, the particle is stabilized to be at 
finite, bulk-valued $\phi_{{\rm min},n}$.

The wave functions $\psi_{n}$ of the system with the two exponentials in the effective potential are in general not known in closed form.
Therefore we will approximate them by the wave functions of the harmonic oscillator potential around the minimum in the next section.

It is interesting to Taylor-expand $V_\text{eff}$ for the $n$-th excited state around its minimum. We have that
\begin{align}
	V_\text{eff}(\phi_\text{min}+\delta \phi)\approx V_\text{eff}(\phi_\text{min})+V''_\text{eff}(\phi_\text{min})\delta\phi^2+\mathcal O(\delta \phi^3),
\end{align}
where
\begin{subequations}
\begin{align}
		V_\text{eff}(\phi_\text{min})&= 4^{-\frac{\gamma }{2 \gamma +\alpha_V}} \left(\frac{\alpha_V}{\gamma }+2\right) \left(\frac{\gamma  n^2}{\alpha_V V_0}\right)^{\frac{\alpha_V}{2 \gamma +\alpha_V}},\\
	V''_\text{eff}(\phi_\text{min})&=	\alpha_V V_0 2^{\frac{v}{2 \gamma +\alpha_V}} (2 \gamma +\alpha_V) \left(\frac{\gamma  n^2}{\alpha_V V_0}\right)^{\frac{\alpha_V}{2 \gamma +v}}.
\end{align}
\end{subequations}
Thus, near the minimum, $\delta \phi_m$ approximately experiences a harmonic oscillator potential.

Consider a small perturbation $\delta \phi$ around the minimum of the effective potential. The particle experiences a restorative force proportional to $2V_\text{eff}''(\phi_\text{min})$. It would be interesting if one could view this as generating a mass to the excited state, where the mass is
$	m_n^2=2V''_\text{eff}(\phi_\text{min})$.
	
	\section{Conclusions}

\label{sec:4}

In this paper, we have examined quantum mechanics on moduli spaces. We have split our work into two parts. In the first half, we have explored non-commutativity between operators and studied how the ESC can provide precise asymptotic formulas governing commutation relations between various operators. In the second half of this paper, we have studied square integrable wave functions in the mini-superspace formalism.

\vskip0.3cm\noindent
In the first part of the work, we used the taxonomy relations among the species scale and brane tensions as functions over the moduli space to derive the 
commutation relations for the corresponding operators. At the asymptotic boundaries of moduli space the taxonomy relations are given in terms of specific constants, such that the related
commutation relations are given in terms of c-numbers. However, inside the moduli space one generically expects that the taxonomy relations are not anymore constant but rather given in terms
of field dependent quantities. This means that the measure of
non-commutativity between two operators becomes itself a field
dependent operator. 
We also discussed
commutation relations and a pattern where the degeneracy in moduli
space is lifted by a non-vanishing potential. In this case, we highlighted some interesting applications to cosmology, which we are planning to investigate in more detail in the future.

\vskip0.3cm\noindent
In the second part of the work, we constructed the quantum mechanical
wave functions over the moduli space. We have focused on cases where the moduli space metric is explicitly known. The non-trivial prototype example is the hyperbolic plane ${\mathbb H}^+$ with duality symmetry
$SL(2,{\bf Z})$, which is the exact moduli space of the type IIB
superstring in ten dimensions. Actually, this moduli space is
particularly interesting because it also characterises many string
compactifications. The quantum wave functions on this moduli space
can be used to compute the vacuum expectation values of the position in moduli space
and also expectation values of the species scale and the tower mass
scales. In fact, the eigenfunctions of the associated Laplace operator and the corresponding spectra are well known from
spectral analysis of quantum mechanics in curved spaces, and
are closely related to the theory of automorphic forms, namely the Eisenstein series and the Maa{\ss} forms.  These forms are also relevant for the theory of quantum chaos and are related to the Riemann conjecture
that asserts that all non-trivial zeros of $\zeta (s)$ lie on the
critical line $\Re s=1/2$ (see e.g.~\cite{zagier}). As we discussed, the same line also determines the moduli
space wave function $\psi$ on ${\mathbb H}^+$, which is given in terms
of the Eisenstein series: 
\begin{equation}
\psi_\beta(\tau,\bar\tau)=E_{{1\over
    2}+i\beta}(\tau,\bar\tau)\, .\nonumber
    \end{equation}
    We have shown that this wave function possesses several interesting properties. 
    The continuous part of the spectrum consists of waves  of momenta $\alpha=\sqrt 2\beta$  that travel through the moduli space and are bounded back at its boundaries.
In addition, for a certain infinite set of  discrete eigenvalues
$\beta_i$, the spectrum contains bound states with positive
energies. For those, the vacuum expectation values of the position
operator in moduli space are fixed to specific values. In physical
terms, these states can be viewed as bound states in a geome\-tric
potential, which is due to the non-trivial geometry of the moduli
space. We have also demonstrated that the species scale and the moduli space wave functions can be asymptotically 
related by replacing the wave momenta $\alpha$ by some specific imaginary numbers that are determined by the $\alpha$-vectors from taxonomy rules discussed before.

\vskip0.3cm\noindent
We also discussed the case where a non-trivial scalar potential is turned on. Again we saw that for potentials that classically lead to run-away solutions for the scalar fields, there can be again bound states, for which quantum effects
stabilise the scalar fields around the minimum of an effective potential. The moduli space wave functions are also interesting since they allow for a new 
definition of a distance on moduli
space~\cite{Demulder:2026cfo} without and also in the presence of a potential (see also \cite{Demulder:2023vlo,Mohseni:2024njl,Demulder:2024glx,Raml:2025yrb} for previous attempts to define distance in field spaces with potential).

\vskip0.3cm\noindent
The  moduli space quantum mechanics describes a way to quantise  operators related to the internal degrees of freedom of an effective gravity theory.
However eventually one is  aiming to quantise also the effective gravity theory, i.e. to quantise the metric of the $d$-dimensional EFT.\footnote{It would be phenomenologically interesting to incorporate effects from other fields, such as spinors, in the resulting moduli-space quantum mechanics.}
For example, this would involve coupling the mini superspace quantum wave function of the space-time metric to the moduli space wave function of our work in an appropriate way.
Since the excited states have positive energy, space-time de Sitter space would possibly then arise from the internal excited  bound states, which we have constructed in our paper.
This way to obtain de Sitter as an excited states in moduli space quantum mechanics is quite along the lines of previous proposals about excited de Sitter states in quantum gravity, as made in 
\cite{Dvali:2014gua,Dvali:2017eba,Dvali:2018fqu,Dvali:2018jhn}.
Then, using these bound states,  one can attempt to investigate the lifetimes of the exited de Sitter states by using for example a domain wall approach.

\section*{Acknowledgements}

We thank 
Severin  L\"ust for a lot of important input and remarks and we also thank
Christian Aoufia, Ivano Basile, Niccolo Cribiori, Bernardo Fraiman, Carmine Montella, Miguel Montero,  Slava Mukhanov, Vinicius Nevoa, Antonia Paraskevopoulou, Sanjay Raman, Thomas Raml, Alex Stewart, Cumrun Vafa, and Ron Zagier for very useful discussions. 
The work of L.A.A. is supported by the U.S. National Science
Foundation (NSF Grant PHY-2412679), he 
 extends his appreciation to the Harvard Swampland Initiative for their warm hospitality and for providing a stimulating environment for productive discussions.
The work of D.L. is supported 
 by the German-Israel-Project (DIP)
on Holography and the Swampland.

\section*{Appendix A: Cosmic expansion}

The deceleration parameter $q$ is a dimensionless quantity that
measures the acceleration or deceleration of the cosmic expansion. It
is defined in terms of the scale factor $a$ and its time derivatives
\begin{equation}
  q \equiv - \frac{\ddot a a}{\dot a^2} \,,
\end{equation}
or more commonly, as the negative ratio of the acceleration to the expansion rate squared
\begin{equation}
  q = - \frac{\ddot a}{a H^2} \, . 
\label{qH}
\end{equation}
Since
\begin{equation}
  \dot H = \frac{\ddot a}{a} -\left(\frac{ \dot a}{a} \right)^2 \,,
\end{equation}
(\ref{qH}) can be recast as (\ref{qdef}).

In observational cosmology, it is common to relate $H$ and $q$ through
the redshift $z$. Indeed, the evolution of the
Hubble parameter can be calculated by integrating the deceleration
parameter over redshift
\begin{equation}
  H(z) = H_0 \exp \left \{\int_0^z \left[1 + q (\xi)\right] \ d \log (1 +\xi) \right\} \, ,
\end{equation}  
with $1 +z = 1/a$. It is instructive to charaterize the different $q$ regimes:
\begin{enumerate}[noitemsep,topsep=0pt]
\item If $q >0$, then $\ddot a < 0$ and $\dot H$ is negative, meaning the expansion is slowing down due to gravity. The
minus sign in the definition makes $q$ positive, hence the name deceleration parameter.
\item If $q<0$, then $\ddot a >0$ and $\dot H$ is positive relative to
  $H^2$ such that $-1 - \dot H/H^2 <0$, 
  indicating that the expansion is speeding up. 
  \begin{enumerate}[noitemsep,topsep=0pt]
    \item If $q =-1$ then $d(\log H)/d\tau = 0$, meaning $H$ is constant
  and the expansion rate is pure exponential; i.e. dS.
\item If $q< -1$, then $d(\log H)/d\tau > 0$,
  meaning $H$ increases and the expansion rate is super-exponential.
  \end{enumerate}
\item  If $q=0$ the expansion rate is constant.
\end{enumerate}

Now, we consider the flat, homogeneous, and isotropic expansion of standard 4D cosmology. The first Friedmann
equation relates the Hubble parameter to the total energy density, 
\begin{equation}
  H^2 = \frac{\rho_{\rm tot}}{3 M_p} \, ,
\end{equation}
where $M_p$ is the reduced Planck mass. We further assume that the energy density is dominated by a
quintessence field
\begin{equation}
  \rho_{\rm tot} \sim \rho_\phi = \frac{1}{2} \dot \phi^2 + V(\phi) \,
  .
\end{equation}
For the scalar field to act as dark energy and cause accelerated
expansion, the field must evolve slowly. This is characterized by the
slow-roll condition, where the kinetic energy density is much smaller
than the potential energy $\dot \phi^2 \ll V(\phi)$. Under this
condition the total energy of the field is dominated by the potential
$\rho_\phi \approx V(\phi)$, and so the Hubble parameter is directly
proportional to the square root of the potential as given in (\ref{HV}). This relationship reflects that the expansion rate is
determined by the nearly constant vacuum-like energy of the scalar
field's potential.

The deceleration parameter is directly linked to
the energy content's pressure-density ratio,
\begin{equation}
q = \frac{1}{2} (1 + 3 \sum w_i \Omega_i) \,,  
\end{equation}
via Friedmann's acceleration equation
\begin{equation}
  \frac{\ddot a}{a} = - \frac{1}{6 M_p} \sum (\rho_i + 3 p_i) \, ,
\label{Fried_acc}
\end{equation}
where $w_i = p_i/\rho_i$ is the equation of state for component $i$
(non-relativistic matter, radiation, dark energy), $\Omega_i =
\rho_i/\rho_{\rm c}$ is its density parameter normalized to the
critical density $\rho_{\rm c}$. Thus, (\ref{Fried_acc}) shows
that $q$ depends on the total effective pressure (related to $\sum w_i
\Omega_i$), linking the observed cosmic deceleration/acceleration to
the nature of dark energy or other components.

If the energy density is dominated by the scalar field, the deceleration parameter can be expressed as
\begin{equation}
  q = \frac{1}{2} ( 1 + 3 w_\phi) \, ,
\end{equation}
with $w_\phi = p_\phi/\rho_\phi$ and 
\begin{equation}
p_\phi = \frac{1}{2} \dot \phi^2 + V(\phi) \, .
\end{equation} 

We can thus examine the impact of $V(\phi)$ on the cosmic expansion:
\begin{itemize}[noitemsep,topsep=0pt]
\item If $V(\phi) \gg \dot \phi^2$, we have the slow-roll condition
  with $w_\phi \alt -1$. This results in
  $-1<q<0$, indicating accelerated expansion.
\item If $\dot \phi^2 \gg V(\phi)$, then $w_\phi$ approaches $+1$
  leading to $q>0$ and decelerated expansion. Particularly, for $w_\phi =1/3$, the
  expansion is dominated by radiation, yielding $q=1$.
\item The universe transitions from deceleration to acceleration when
  $w_\phi < -1/3$. This occurs when $q=0$.
\item If $\dot \phi^2 =0$, then $V(\phi) =$ constant, $q=-1$, and the expansion rate is pure exponential. 
\item If $q<-1$, then $\dot \phi^2 < 0$, the expansion rate is
  super-exponential, yielding violation of
  unitarity. 
\end{itemize}
All in all the physical interval where $V(\phi)$ dominates the
expansion, relevant to (\ref{ansscommutator2}), is
$-1 < q \leq 0$.

\end{document}